\documentclass[prd,aps,apl,reprint,superscriptaddress,nofootinbib,preprintnumbers]{revtex4-1}

\usepackage{graphicx}
\usepackage{amsmath}
\usepackage{dcolumn}
\usepackage{bm}

\begin{document}

\preprint{ICRR-Report-640-2012-29,\ IPMU12-0217}

\title{
Opening the window to the cogenesis with Affleck-Dine mechanism 
in gravity mediation
}

\author{Ayuki Kamada}
\affiliation{Kavli Institute for the Physics and Mathematics of the Universe, 
     University of Tokyo, Kashiwa, Chiba 277-8582, Japan}
     
\author{Masahiro Kawasaki}
\affiliation{Kavli Institute for the Physics and Mathematics of the Universe, 
     University of Tokyo, Kashiwa, Chiba 277-8582, Japan}
\affiliation{Institute for Cosmic Ray Research, 
     University of Tokyo, Kashiwa, Chiba 277-8582, Japan}

\author{Masaki Yamada}
\affiliation{Institute for Cosmic Ray Research, 
     University of Tokyo, Kashiwa, Chiba 277-8582, Japan}

\date{\today}

\begin{abstract}
The observed baryon and dark matter densities are equal up to a factor of 5. 
This observation indicates that the baryon asymmetry and dark matter have the same origin. 
The Affleck-Dine baryogenesis is one of the most promising mechanisms in this context. 
Q balls, which are often formed in the early Universe associated with 
the Affleck-Dine baryogenesis, decay both into supersymmetric particles 
and into quarks. 
Recently, it was pointed out that annihilation of squarks into quarks 
gives a dominant contribution to the Q-ball decay rate and 
the branching ratio of Q-ball decay into supersymmetric particles changes 
from the previous estimate. 
In this paper, the scenario of baryon and dark matter cogenesis from Q ball 
in gravity mediation is revisited  in respect of the improved Q-ball decay rates. 
It is found that the successful cogenesis takes place when a wino with mass $0.4-1$~TeV
is dark matter. 
\end{abstract}

\pacs{98.80.Cq, 95.35.+d, 12.60.Jv}

\maketitle

\section{\label{sec1}Introduction}

The existence of the baryon asymmetry and the dark matter
is a long standing challenge in cosmology and particle physics. 
In supersymmetric (SUSY) extensions of the Standard Model (SM), 
the lightest SUSY particle (LSP) is a good candidate for dark matter 
if the R-parity is conserved. 
Furthermore, 
the Affleck-Dine mechanism can provide the baryon asymmetry~\cite{AD, DRT}. 
In the gravity-mediated SUSY breaking model, the Affleck-Dine mechanism often predicts 
the formation of Q balls in the early universe~\cite{KuSh, EnMc, KK1,KK2,KK3}. 
The Q ball is a spherical condensate of scalar fields. 
It generally consists of squarks and sleptons, and 
eventually decays both into quarks and into SUSY particles 
before the Big Bang Nucleosynthesis (BBN), and the observed baryon asymmetry is released. 
Through the cascade decays, the SUSY particles produced by the Q-ball decay turn into LSPs, 
which can account for the dark matter in the Universe. 
In this case, the baryon asymmetry and dark matter have the same origin and 
the resultant ratio of baryon to dark matter can be $O(1)$ 
naturally~\cite{WMAP, EnMc, Kap, FH1, FH2, FY}. 

When we consider the case that the pair annihilation of the LSPs is ineffective 
and assume that the Affleck-Dine field $\phi$ takes a circular orbit in the complex $\phi$ plane, 
the resultant ratio of baryon to dark matter from the Q-ball decay is
related only with the mass of the LSP and the branching ratio of the Q-ball decay into 
baryons and SUSY particles. 
In the previous works, the branching ratio of the Q-ball decay into SUSY particles 
is believed to be comparable with that into quarks~\cite{EnMc, Kap, FH1, FH2, FY, ShKu, DoMc}. 
In this case, the mass of dark matter should be $O(1)$GeV.\footnote{
If we consider the case that the pair annihilation is effective, 
the resultant LSP density is determined by the mass of LSP, the pair annihilation rate of LSP 
and the decay temperature of the Q ball~\cite{FH1, FH2, FY}. 
Thus, the branching ratios of the Q-ball decay do not affect the ratio of the baryon to LSPs.}
However, it was pointed out that 
the many body processes like the squark annihilation may be dominant and then 
the branching ratio may change drastically~\cite{KK2011}. 
In this letter, we reexamine the branching ratio into SUSY particles 
in respect of the many body process. 

Since the effective mass of the squark inside the Q ball is smaller than that of the free
squark, the Q ball cannot decay into squarks. 
We assume that the Q ball is kinematically allowed to decay into binos, winos (LSPs), 
and SM particles. 
When the Q-ball decay rate is saturated due to the Pauli exclusion principle~\cite{evap}, 
the branching ratio is 
determined only by the number of degrees of freedom in the final state. 
Finally, we show that the branching ratio into SUSY particles can be $O(0.01)$. 
By using this branching ratio, we provide a successful scenario of the 
baryon and dark matter cogenesis through the Q-ball decay, 
and show that the wino LSP with mass of $0.4-1$~TeV 
can naturally explains the observed baryon to dark matter ratio 
in the case that the pair annihilation of the LSPs is ineffective. 

This letter is organized as follows. 
In Sec.~\ref{sec2}, we briefly review the property of Q balls 
in gravity mediation. 
In Sec.~\ref{sec3}, 
first we compare the saturated decay and annihilations and then 
derive the branching ratios. 
In Sec.~\ref{sec4}, 
we discuss the thermal history in our scenario. 
Sec.~\ref{conclusion} is devoted to the conclusion.

\section{\label{sec2}Q ball properties in gravity mediation}
In SUSY extensions of the standard model, there are many flat directions 
in the scalar potential. 
The flat directions are lifted by the SUSY breaking effect, 
and we can take the following potential for the flat direction to see the property of the Q ball 
in gravity mediation: 
\begin{equation}
  V = m^2_\phi \vert \phi \vert^2 
  \left( 1+ K \log \frac{ \vert \phi \vert^2}{M^2_{\rm P}} \right), 
\end{equation}
where $m_\phi$ is the mass of the flat direction and $M_{\rm P}$ is the reduced 
Planck mass($\simeq 2.4\times 10^{18}$~GeV). 
In gravity mediation, $m_s$ 
is the same order of the gravitino mass $m_{3/2}$. 
The second term in the parenthesis comes from the one-loop radiative corrections, 
and typically $ \vert K \vert  \sim$0.01-0.1. 
In many cases, the gluino loops have dominant contributions to the radiative corrections and lead to $K<0$, 
and then there exists a Q-ball solution~\cite{EnMc, Coleman}. 
The energy of the Q ball $M_Q$, the radius $R$, the rotation speed of the field $\omega_0$, 
and the field amplitude at the center of the Q ball $\phi_0$ are given by 
\begin{eqnarray}
  M_Q & \simeq  & m_\phi(\phi_0) Q, \\
  R  & \simeq & \frac{1}{ \vert K \vert^{1/2} m_\phi(\phi_0)}, \label{gravproperty} \\
  \omega_0 & \simeq &  m_\phi(\phi_0), \\
  \phi_0 & \simeq & (2 \pi^{3/2} )^{-1/2} \vert K \vert^{3/4} 
  m_\phi(\phi_0) Q^{1/2}, \label{phi}
\end{eqnarray}
where $m_\phi(\phi_0)$ is the mass defined at the energy scale $\phi_0$. 
The rotation speed $\omega_0$ has a further important meaning as 
$\omega_0 = \text{d}M_Q /\text{d}Q$; i.e., 
the Q-ball energy per unit charge. 

As discussed in detail in Sec.~\ref{sec4}, 
the decay temperature of Q balls should be sufficiently suppressed 
for the pair annihilation of LSPs to be ineffective. 
This indicates that the charge of Q balls should be 
$Q \gtrsim 10^{26}$ 
and thus the magnitude of the scalar field is $\phi_0 \gtrsim 10^{13} m_\phi(\phi_0)$. 
At this energy scale, the mass of the flat direction $m_\phi(\phi_0)$ 
is lower than the mass of squarks at the electro-weak scale due to $K<0$, 
and the Q ball cannot decay into squarks.

\section{\label{sec3}Q-ball decay rates into bino-wino, and quarks}

The fermion production rates from the Q ball have upper bounds 
due to the Pauli exclusion principle~\cite{evap}. 
The upper bound of the each massless fermion flux $\bm{j}$ from the Q-ball surface is calculated as 
\begin{eqnarray}
 \bm{n\cdot j} &\lesssim & 2 \int \frac{\text{d}^3 k}{(2 \pi)^3} 
 \theta \left( \omega_0/2-|\bm{k}| \right) 
 \theta(\bm{k \cdot n}) \bm{\hat{k}\cdot n}, \\
 &= & \frac{2}{8 \pi^2}  \int^{\omega_0/2}_0 k^2 \text{d}k 
 = \frac{\omega_0^3}{96 \pi^2}, \label{saturate}
\end{eqnarray}
where $\bm{n}$ is the outward-pointing normal. 
We double the flux and take 
the upper limit of integration as $\omega_0/2$, because 
one of the decay products has energy less than $\omega_0/2$. 
We obtain the upper bound for the production rate from the Q ball 
by multiplying Eq.~(\ref{saturate}) by the area of the Q-ball surface $4 \pi R^2$. 
The decay rate is saturated when $g \phi_0 > \omega_0$ for the interaction 
$g \phi \xi \eta$ ($\xi$, $\eta$: massless fermions). 
The condition $g\phi_0 > \omega_0$ is 
almost always satisfied due to the large $Q$ value (see Eq.~(\ref{phi})). 

In the case of the massive fermion $\chi$, the upper bound of the flux is lower than Eq.~(\ref{saturate}). 
We consider the process of squark $\to$ quark $+\chi$, and treat the quark as a massless particle. 
The fermion $\chi$ can obtain the energy in the range of $[m_{\chi}, \omega_0]$, 
and the quark obtain the energy in the range of $[0, \omega_0-m_{\chi}]$. 
Taking this into account, we just change the integral of Eq.~(\ref{saturate}) as 
\begin{equation}
  \frac{1}{8 \pi^2}  \int^{\omega_0 - m_{\chi}}_0 k^2 \text{d}k, 
\end{equation}
for $\omega_0 > m_{\chi} > \omega_0/2$, and as 
\begin{equation}
  \frac{1}{8 \pi^2} \left[ \int^{\omega_0/2}_0 k^2 \text{d}k 
  + \int^{\omega_0/2}_{m_{\chi}} k^2 \text{d}k \right], 
\end{equation}
for $m_{\chi}  < \omega_0/2$. 
Thus, the $\chi$ flux is given by 
\begin{eqnarray}
  \bm{n\cdot j}_{\chi} & \simeq & \frac{\omega_0^3}{96 \pi^2} 
  \times f(m_{\chi}/\omega_0), \label{bino massive}\\[0.6em]
  f(x) & \equiv &
  \left\{ 
  \begin{array}{ll}
      4 (1-x)^3 \quad \text{ for } 1/2 < x < 1, \\[0.4em]
      4 [(1/2)^3 + ((1/2)^3-x^3)]  \quad \text{ for }x  < 1/2. 
  \end{array}\right.
  \label{funcx}
\end{eqnarray}

Q balls can also decay into quarks via heavy gluino/higgsino exchange $\phi \phi \to q q$. 
This reaction rate is also saturated by the Pauli exclusion principle. 
The detailed discussion is given in Ref.~\cite{KK3, KY}. 
The saturated flux is Eq.~(\ref{saturate}) 
with $\omega_0$ replaced by $2\omega_0$, 
which is the total energy available in this process. 
Thus, we obtain the each quark flux as 
\begin{eqnarray}
  \bm{n\cdot j}_{\text{quark}} \simeq \frac{(2\omega_0)^3}{96 \pi^2}. 
  \label{scat}
\end{eqnarray}
This is larger than Eq.~(\ref{saturate}) by a factor of 8. 
Notice that 
this flux is valid only for $M > \omega_0$, where $M$ is the gluino/higgsino mass, 
and we assume it in this letter. 
In Appendix, we show $N (\ge 3)$ body processes are not saturated and negligible.

Now, let us compare the branching ratios of the Q-ball decay into binos, winos, and quarks. 
The bino or wino production rate is given by Eq.~(\ref{bino massive}), 
while the quark production rate is given by Eq.~(\ref{scat}). 
Here we should note that since the saturated production rate is determined by the Pauli exclusion principle, 
the total quark production rate is Eq.~(\ref{scat}) times the number of 
degrees of freedom for quarks produced in the decay. 
We can count it once we specify the flat direction. 
Hereafter, we consider the flat direction 
$\bar{u}^a_{i} \bar{d}^b_{j} \bar{d}^c_{k} \epsilon_{abc}$ ($j \neq k$), 
where $a$, $b$, and $c$ are the color indices and $i$, $j$, and $k$ are the family indices. 
The Q ball can decay into all right handed quarks via gluino exchange and 
into all left handed quarks via higgsino exchange, 
because the flat direction contains all colors and, in general, all families. 
(Even if the flat direction does not contain all families, it can decay into 
all families through flavor mixings.)
The $U(1)_Y$ charge conservation allows one up-type quarks for each two down-type quarks. 
The Q ball cannot directly decay into winos 
because the $\bar{u} \bar{d} \bar{d}$ flat direction has no tree-level interaction with winos. 
However, winos are produced via subsequent decays of binos if the LSP is wino.
In this letter we consider winos as the LSPs. 

We conclude that the total decay rate of the Q ball and 
the branching ratios of the decay into quarks and bino are calculated as 
\begin{eqnarray}
  \sum_{\text{all}} \frac{\text{d}N}{\text{d}t} &\simeq &
  \left[ 8 \times 36 \times \frac{3}{4} 
  + f\left(\frac{m_{\tilde{b}}}{\omega_0} \right) \right] 
  \frac{R^2 \omega_0^3}{24 \pi},   
  \label{total decay rate}\\
  B_{\text{quarks}} 
  &\simeq & \frac{8 \times 36 \times 3/4}{8 \times 36 \times 3/4
  + f(m_{\tilde{b}}/\omega_0) }, \\
  B_{\text{bino}} &\simeq & 
  \frac{f(m_{\tilde{b}}/\omega_0)}{8 \times 36 \times 3/4 
  + f(m_{\tilde{b}}/\omega_0) }. 
  \label{ratio}
\end{eqnarray}
In Eq.(\ref{total decay rate}) 
the factor $36$ comes from the degrees of freedom for 
colors (3), flavors (6) and chiralities (2), and 
the factor $3/4$ comes from the $U(1)_Y$ charge conservation.
We do not include the quarks from the process of squark $\to$ quark + bino, 
because the quarks production rates are determined by the Pauli exclusion principle 
and the phase space of the quarks produced by the squark decay is 
a subset of that of the quarks produced by the squark annihilation.
The binos eventually decay into winos (LSPs). 
Note that the thermal relics of winos do not overclose the Universe.\footnote{%
On the other hand, if binos with mass larger than a few $\times 100$~GeV are the LSPs, 
their relic density may overcloses the Universe .}
%

\section{Cogenesis in gravity mediation}
\label{sec4}

In this section, we show that our scenario of the baryon and dark matter cogenesis
works well. 
We consider the Affleck-Dine mechanism using the flat direction without non-renormalizable superpotential, 
because our scenario requires Q balls with $Q \gtrsim 10^{26}$ for the pair annihilation to be ineffective. 
The scalar potential for the flat direction is typically written as
\begin{eqnarray}
  V(\phi) &=& (m_\phi^2 -c_H H^2) |\phi|^2 
  +\frac{m_{3/2}^2}{M_*^{n-2}} (a_m \phi^n + h.c.)  
  \nonumber\\
  &+& \frac{H^2}{M_*^{n-2}}\left( a_H \phi^{n} + h.c. \right)+ \ldots, 
  \label{eq:potential}
\end{eqnarray}
where $M_*$ is a cut-off scale and $\ldots$ denotes higher order Planck-suppressed terms. 
The terms proportional to $H^2$ are 
induced via the interaction with the inflaton, and 
$c_H$, $a_m$, and $a_H$ are $O(1)$ constants. Here we assume $c_H>0$. 
Owing to the Hubble induced terms and higher order Planck-suppressed terms, 
the flat direction has a large expectation value during inflation $\phi \simeq M_*$, 
and then begins to oscillate and rotate around $\phi=0$ when $H \simeq m_\phi$. 
Soon after the oscillation, Q balls are formed.
Here we assume that the second term in Eq.(\ref{eq:potential}) which kicks $\phi$ 
in the phase direction is large enough for $\phi$ to take a circular orbit.
In this case anti-Q balls are not produced, which leads to the simple relation 
between baryon and  dark matter densities.\footnote{%
When both Q balls and anti-Q balls are formed, cogenesis requires much 
smaller LSP mass.  Such small LSP mass is realized in gauge mediated SUSY breaking
where a gravitino with mass $\lesssim 1$~GeV is the LSP~\cite{Kasuya:2012mh}.}
The charge of the Q ball is determined by  
\begin{equation}
  Q  \sim \beta \left( \frac{M_*}{m_\phi} \right)^{2} 
  \sim 3 \times 10^{28}  \left( \frac{M_*}{M_\text{P}} \right)^{2}  
  \left( \frac{2\text{ TeV}}{m_\phi(\phi_0)} \right)^{2}, 
\end{equation}
where $\beta=2\times 10^{-2}$~\cite{Qgrav}. 

The Q-ball decay temperature is estimated as 
\begin{eqnarray}
  T_{\text{d}} 
  & =& \left( \frac{90}{4 \pi^2 g_*} \right)^{1/4} \sqrt{\Gamma_Q M_\text{P}}, 
  \nonumber\\
  &\simeq& 10\text{ MeV} \left( \frac{m_\phi(\phi_0)}{2\text{ TeV}} \right)^{1/2} 
  \left( \frac{10^{28}}{Q} \right)^{1/2}, \label{decay temp}
\end{eqnarray} 
where $g_*$ is the effective relativistic degrees of freedom at the decay time, 
and $\Gamma_Q = (1/Q) \sum_\text{all} \text{d}N/\text{d}t $ is the decay rate of the Q ball. 
In the second line of Eq.~(\ref{decay temp}), 
we set $g_* = 10.75$ and 
$\sum_\text{all} \text{d}N/\text{d}t \sim 200 \times R^2 \omega_0^3 /24 \pi Q$.  
We find that the Q ball decays before the BBN 
but after the sphaleron process freezes out~\cite{sphaleron}.
Winos produced from the Q-ball decay do not annihilate 
when the following condition is satisfied:
\begin{eqnarray}
  Y_{\tilde w}^{(NT)} &\ll & \sqrt{\frac{45}{8\pi^2 g_*}} 
  \frac{1}{\langle{\sigma v} \rangle M_\text{P} T_d}, \\
  &\simeq & 1.1 \times 10^{-10} \times 
  \left(\frac{10^{-24}{\rm cm^3/s}}{\langle{\sigma v}\rangle }\right)
  \left(\frac{10\,{\rm MeV}}{T_d}\right)\ .
  \label{wino constraint}
\end{eqnarray}

As mentioned above, we consider the flat direction without non-renormalizable superpotential. 
In this case, the Q balls dominate the Universe soon after inflation, 
and the baryon-to-entropy ratio is given by 
\begin{equation}
  Y_{b} \simeq 10^{-10}\left(\frac{T_{\text{d}}}{10\text{ MeV}}\right)
  \left(\frac{2\text{ TeV}}{m_{\phi}(\phi_0)}\right)\left(\frac{10^4}{\Delta}\right), 
\end{equation}
where we include the dilution factor $\Delta$. 
There is some mechanism to produce entropy after the reheating of the inflation, such as 
thermal inflation~\cite{thermal inf} and domain wall decay~\cite{domain wall}. 
We do not specify the dilution mechanism and assume that the baryon asymmetry 
produced from the Q-ball decay 
is consistent with the observation. 
A dilution mechanism may also dilute the undesirable relics such as thermal relic of the stable bino.
Thus, the successful bino LSP scenario may be realized as a simple extension of the present wino LSP scenario.
However, most of the dilution mechanisms produce SUSY particles at the same time.
This is a reason why we focus on the wino LSP scenario.

\begin{figure}[t]
  \includegraphics[width=85mm]{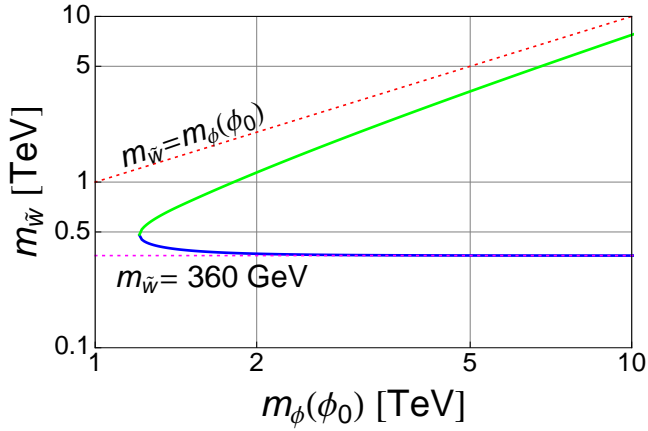}
  \caption{Solutions of Eq.~(\ref{wino eq}) for the case of $m_{\tilde{w}} = m_{\tilde{b}}$ 
  as a function of $m_\phi(\phi_0)$. 
  There is no solution for $m_\phi(\phi_0) \lesssim 1.2$~TeV 
  and are two independent solutions for $m_\phi(\phi_0) \gtrsim 1.2$~TeV (green and blue lines). 
  The red and magenta dotted lines show the two asymptotic solutions 
  $m_{\tilde{w}} = m_\phi(\phi_0)$ and $m_{\tilde{w}} = 360$~GeV, respectively. }
  \label{fig1}
\end{figure}

In the case of the wino LSP, 
the thermal relic abundance can be ignored for $m_{\tilde{w}} \ll 1\text{ TeV}$~\cite{winoLSP}. 
The baryon-to-dark matter ratio is determined only by the Q-ball decay: 
\begin{eqnarray}
  5 \simeq \frac{\Omega_{\text{DM}}}{\Omega_b} 
  = \frac{3 m_{\tilde{w}}}{m_N} \frac{B_{\text{bino}}}{B_{\text{quarks}}}. 
  \label{BDMratio} 
\end{eqnarray}
From Eqs.~(\ref{funcx}), (\ref{ratio}), and (\ref{BDMratio}),  
the bino and wino masses are related with each other by the following equation: 
\begin{equation}
  \frac{360\text{ GeV}}{m_{\tilde{w}}} = f\left( \frac{m_{\tilde{b}}}{\omega_0}\right), 
  \label{wino eq}
\end{equation}
where $f(x)$ is defined as Eq.~(\ref{funcx}). 
The results are shown in Fig.~\ref{fig1} for the case of $m_{\tilde{w}} = m_{\tilde{b}}$. 
There are two solutions when $\omega_0 \gtrsim 1.2$~TeV. 
In the limit of $\omega_0 \to \infty$, two solutions are approximated to 
$360$~GeV and $\omega_0$. 
The resultant wino abundance is given by 
\begin{equation}
  Y_{\tilde w}^{(NT)} \simeq 1.1\times10^{-12}\frac{360\text{ GeV}}{m_{\tilde w}}. 
\end{equation}
From this and Eq.~(\ref{wino constraint}), 
we can check that the winos with mass of $0.4-1$~TeV
do not annihilate 
when $T_{\text{d}} \lesssim 100$~MeV ($Q \gtrsim 10^{26}$). 
Indirect detection experiments constrain the wino mass
as $m_{\tilde{w}} \gtrsim 300$ GeV~\cite{FermiLAT, PAMELA, puregrav}. 
The above predicted wino mass satisfies this constraint. 

\section{\label{conclusion}conclusions}

We have reinvestigated the baryon and dark matter cogenesis through Q-ball decay 
into quarks and SUSY particles 
by taking into account the squark annihilation process inside the Q ball.  
The branching ratio of the Q-ball decay into quarks is enhanced 
by the number of degrees of freedom 
for quarks produced in the decay. 
We have assumed that the Q ball can decay into binos, winos, and  SM
particles kinematically, and considered the wino as LSP.  
In this case, we show that the branching into binos can be $O(0.01)$ 
for the $\bar{u}\bar{d}\bar{d}$ flat direction
and predict that the dark matter is the wino with mass of $0.4-1$~TeV. 

\appendix*
\section{Q-ball decay rates through the $N \ge 3$ body scattering processes}

Not only the decay process but also the $N$ body scattering processes can occur in the Q ball. 
The rate of the charge emission from the Q ball through the $N$ body scattering process 
can be roughly estimated as 
\begin{eqnarray}
  \left( \frac{\text{d}N}{\text{d}t} \right)_N  & \sim & 
  Q \times n_\phi^{N-1} \times \Gamma_N, \\
  \Gamma_N  & = & \int \text{dLips} |\mathcal{M}|^2 
  \prod_{\text{initial}} \frac{1}{2E_i}, \\
  \text{dLips} & \equiv & (2\pi)^4 \delta \left( \sum_{\text{all}} p_j \right) 
  \prod_{\text{final}} \frac{\text{d}^3k_i}{(2 \pi)^3 2 E_i}, 
\end{eqnarray}
where $n_\phi \sim \omega_0 \phi_0^2$ is the squark number density in the Q ball. 
Let us show that the rates of the $N$ body scattering processes are 
not saturated for $N \ge 3$.

\begin{figure}[b]
  \begin{tabular}{l}
    \resizebox{35mm}{!}{\includegraphics{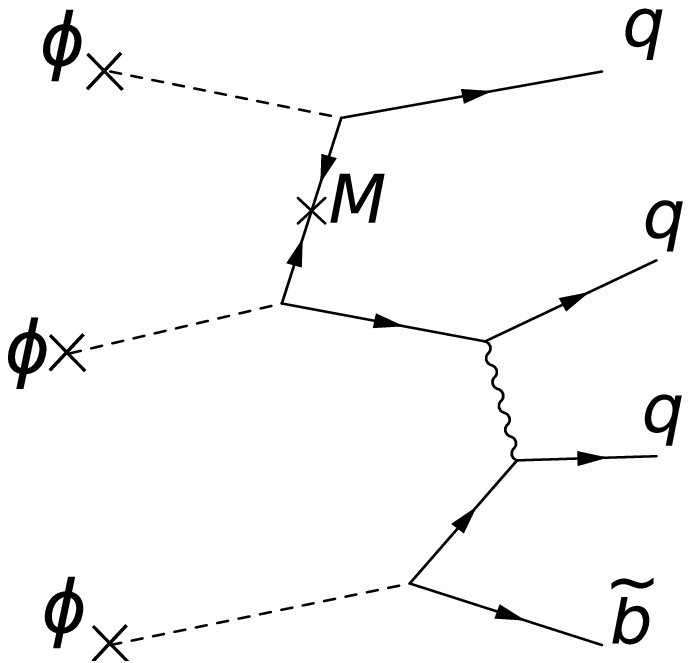}} \qquad
    \resizebox{35mm}{!}{\includegraphics{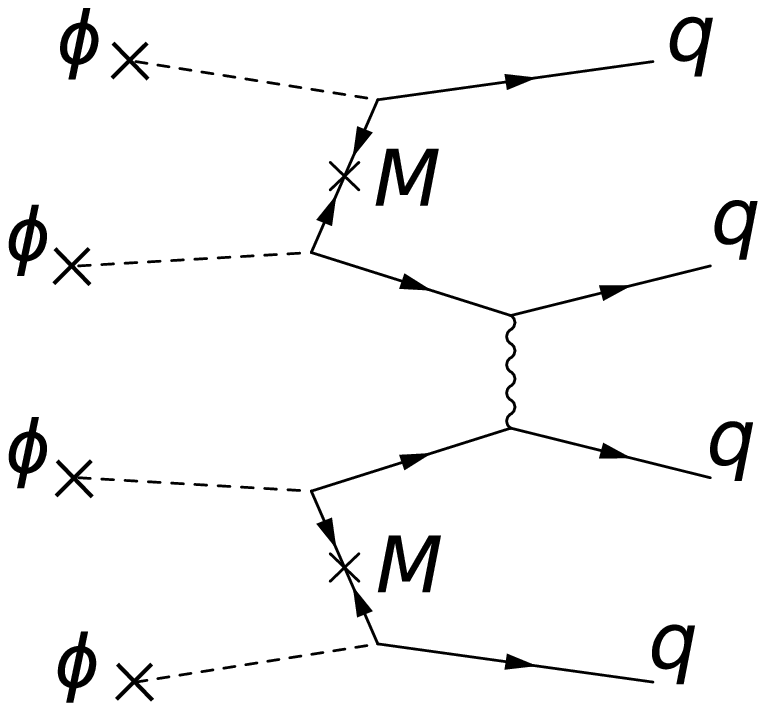}} 
  \end{tabular}
  \caption{Examples of the diagrams for the $N$ body scattering processes. }
  \label{fig2}
\end{figure}

The mass of the field interacting with the Q ball is $O(\phi_0)$, 
but the typical interaction energy is $O(\omega_0)$. 
Thus, we can estimate the rates of the $N$ body scattering processes 
in the leading order of $\omega_0 / \phi_0 \sim Q^{-1/2}$. 
The number of particles in the final state should be minimized in the leading order as 
\begin{equation}
  N_{\text{ext}} = 
  \left\{ 
  \begin{array}{ll} 
     N, \qquad ~~~& \text{N: even}\\[0.4em] 
     N+1, \qquad  & \text{N: odd}. 
  \end{array}
  \right.
\end{equation}
Then, the number of fermion propagators can be counted as 
\begin{equation}
  N_{\text{prop}} = 
  \left\{ 
  \begin{array}{ll} 
     3N/2-2, \qquad ~~~& \text{N: even}\\[0.4em] 
     3N/2-3/2, \qquad & \text{N: odd}. 
  \end{array}
  \right.
  \label{Nprop}
\end{equation}
However, as shown in Fig.~\ref{fig2}, 
there should be a factor of $M$ from the chirality flip, where $M$ is the Majorana gluino mass 
or the higgsino mass, and we assume $\omega_0 < M \ll \phi_0$. 
The number of mass insertions is 
\begin{equation}
  N_{\text{mass}} = 
  \left\{ 
  \begin{array}{ll} 
      N/2, \qquad ~~~& \text{N: even}\\[0.4em] 
      (N-1)/2,  \qquad & \text{N: odd}. 
  \end{array}
  \right.
  \label{Nmass}
\end{equation}
The gauge boson is massless if it has no tree level interaction with the Q ball. 
Hereafter, we conservatively take the gauge boson as a massless field. 
Thus, from Eqs.~(\ref{Nprop}) and (\ref{Nmass}), we can estimate $|\mathcal{M}|^2$ as 
\begin{equation}
  |\mathcal{M}|^2 \sim 
  \left\{ 
  \begin{array}{ll} 
     \phi_0^{8-6N} M^{N} \omega_0^{N}, ~~~& \text{N: even}\\[0.5em]
     \phi_0^{6-6N} M^{N-1} \omega_0^{N+1}, &  \text{N: odd} 
  \end{array}
  \right.
\end{equation}
Here we determine the $\omega_0$ dependence from dimensional analysis. 
On the other hand, the kinematics is determined only by $\omega_0$. 
We conclude that the charge emission rates from Q ball 
through the $N$ body scattering process can be estimated as 
\begin{eqnarray}
   \left( \frac{dN}{dt} \right)_N &\sim& Q (\omega_0 \phi_0^2)^{N-1} \Gamma_N, 
   \nonumber\\
    &\sim& 
    \left\{ 
    \begin{array}{ll}
      \omega_0 Q^{4-2N} (M/\omega_0)^N, ~~~& \text{N: even}\\[0.4em]
      \omega_0 Q^{3-2N} (M/\omega_0)^{N-1}, & \text{N: odd}
    \end{array}
    \right. 
\end{eqnarray}
where we have used $Q\sim \phi_0^2/\omega_0^2$. 
We should compare this with the saturated emission rate from the Q ball 
$(\text{d}N/\text{d}t)_{\text{sat}} \sim \omega_0$ (see Eqs.~(\ref{gravproperty}) 
and (\ref{total decay rate})) 
and find that the rate is not saturated for $N \ge 3$.

\begin{acknowledgments}
This work is supported in part by JSPS Research Fellowship for Young Scientists (A.K.), 
by Grant-in-Aid for Scientific Research from
the Ministry of Education, Science, Sports, and Culture (MEXT), Japan,
No.\ 14102004 (M.K.), No.\ 21111006 (M.K.)
and by World Premier International Research Center Initiative, MEXT, Japan. 
Numerical computation in this work was carried out in part at the 
Yukawa Institute Computer Facility. 
\end{acknowledgments}




\begin{thebibliography}{90}

\bibitem{AD}
  I.~Affleck and M.~Dine,
  Nucl.\ Phys.\  B {\bf 249}, 361 (1985).

\bibitem{DRT} 
  M.~Dine, L.~Randall and S.~D.~Thomas,
  Nucl.\ Phys.\ B {\bf 458}, 291 (1996).

\bibitem{KuSh}
  A.~Kusenko and M.~E.~Shaposhnikov,
  Phys.\ Lett.\  B {\bf 418}, 46 (1998).
  
\bibitem{EnMc}
  K.~Enqvist and J.~McDonald,
  Phys.\ Lett.\  B {\bf 425}, 309 (1998); 
  Nucl.\ Phys.\  B {\bf 538}, 321 (1999).
  
\bibitem{KK1}
  S.~Kasuya and M.~Kawasaki,
  Phys.\ Rev.\  D {\bf 61}, 041301(R) (2000).

\bibitem{KK2}
  S.~Kasuya and M.~Kawasaki,
  Phys.\ Rev.\  D {\bf 62}, 023512 (2000).

\bibitem{KK3}
  S.~Kasuya and M.~Kawasaki,
  Phys.\ Rev.\  D {\bf 64}, 123515 (2001).


\bibitem{WMAP} 
  E.~Komatsu {\it et al.}  [WMAP Collaboration],
  Astrophys.\ J.\ Suppl.\  {\bf 192}, 18 (2011).
  


\bibitem{Kap} 
  D.~B.~Kaplan,
  Phys.\ Rev.\ Lett.\  {\bf 68}, 741 (1992).


\bibitem{FH1} 
  M.~Fujii and K.~Hamaguchi,
  Phys.\ Lett.\ B {\bf 525}, 143 (2002).

\bibitem{FH2} 
  M.~Fujii and K.~Hamaguchi,
  Phys.\ Rev.\ D {\bf 66}, 083501 (2002).
  

\bibitem{FY} 
  M.~Fujii and T.~Yanagida,
  Phys.\ Lett.\ B {\bf 542}, 80 (2002).


\bibitem{ShKu}
  I.~M.~Shoemaker and A.~Kusenko,
  Phys.\ Rev.\  D {\bf 80}, 075021 (2009).
  
\bibitem{DoMc} 
  F.~Doddato and J.~McDonald,
  JCAP {\bf 1106}, 008 (2011).
  
\bibitem{KK2011}
S.~Kasuya and M.~Kawasaki,
  Phys.\ Rev.\ D {\bf 84}, 123528 (2011).

\bibitem{evap}
  A.~G.~Cohen, S.~R.~Coleman, H.~Georgi and A.~Manohar,
  Nucl.\ Phys.\  B {\bf 272}, 301 (1986).
  
  \bibitem{Coleman}
S.~Coleman,
Nucl.\ Phys.\ {\bf B262} (1985) 263.

\bibitem{KY}
  M.~Kawasaki and M.~Yamada,
  arXiv:1209.5781 [hep-ph].
  
\bibitem{Kasuya:2012mh} 
  S.~Kasuya, M.~Kawasaki and M.~Yamada,
  arXiv:1211.4743 [hep-ph].
  
\bibitem{Qgrav} 
  T.~Hiramatsu, M.~Kawasaki and F.~Takahashi,
  JCAP {\bf 1006}, 008 (2010).
    
\bibitem{sphaleron} 
  V.~A.~Kuzmin, V.~A.~Rubakov and M.~E.~Shaposhnikov,
  Phys.\ Lett.\ B {\bf 155}, 36 (1985).
  
  
\bibitem{thermal inf} 
  D.~H.~Lyth and E.~D.~Stewart,
  Phys.\ Rev.\ D {\bf 53}, 1784 (1996); 
  Phys.\ Rev.\ Lett.\  {\bf 75}, 201 (1995). 
  
\bibitem{domain wall} 
  M.~Kawasaki and F.~Takahashi,
  Phys.\ Lett.\ B {\bf 618}, 1 (2005).

\bibitem{winoLSP} 
  J.~Hisano, S.~Matsumoto, M.~Nagai, O.~Saito and M.~Senami,
  Phys.\ Lett.\ B {\bf 646}, 34 (2007).
  
\bibitem{FermiLAT} 
  A.~Charbonnier, C.~Combet, M.~Daniel, S.~Funk, J.~A.~Hinton, D.~Maurin, C.~Power and J.~I.~Read {\it et al.},
  Mon.\ Not.\ Roy.\ Astron.\ Soc.\  {\bf 418}, 1526 (2011).
 
\bibitem{PAMELA} 
  O.~Adriani {\it et al.}  [PAMELA Collaboration],
  Phys.\ Rev.\ Lett.\  {\bf 105}, 121101 (2010).
  
\bibitem{puregrav} 
  M.~Ibe, S.~Matsumoto and T.~T.~Yanagida,
  Phys.\ Rev.\ D {\bf 85}, 095011 (2012).

    
  
\end{thebibliography}
\end{document}